# New CCD Photometric Investigation of High Amplitude δ Scuti Star V2455 Cyg

S. Ostadnezhad, Gh. Forozani and M. Ghanaatian

Department of Physics, Payame Noor University (PNU), P.O. Box 19395-3697, Tehran, Iran; *forozani@pnu.ac.ir*

**Abstract** New *V*-band CCD observations of variable star V2455 Cyg were performed during two nights in September 2017. According to all times of maximum light and new maxima, the O-C curve was analyzed. The period changes of V2455 Cyg were investigated and the rate of increasing period was obtained to be $(1/P)\, dP/dt = 1.99 \times 10^{-7}\, yr^{-1}$. Frequency analysis indicated that V2455 Cyg pulsates with the radial p-mode and the fundamental frequency is $10.61574\, d^{-1}$. Physical parameters of V2455 Cyg at mean temperature, were determined (e.g, $R = 2.52\, R_\odot$ and $M = 1.92\, M_\odot$). The position of this star in H-R diagram confirms that V2455 Cyg is a high amplitude δ Scuti star.

## 1 INTRODUCTION

The pulsating star V2455 Cyg (GSC 03590-01884, TYC 3590-1884-1, HD 204615, SAO 50907, NSV 25610, BD+46 3325, Gaia DR1 1972215241060681728, Gaia DR2 1972215245359762816, 2MASS J21282456+4640308) was observed for the first time by Yoss et al. (1991). They determined the spectral type F2, visual magnitude $V = 8.86\, mag$, absolute magnitude $M_V = 2.2\, mag$, index color $B - V = 0.27$, the distance of 215 pc and space velocity $S = 32\, kms^{-1}$. Piquard (2001) suggested that V2455 Cyg is a SX Phe variable star with the period of $0.094206\, d$ (Wils et al. 2003), while Wils et al. reported that V2455 Cyg is a high amplitude δ Scuti star (HADS). They also improved the ephemeris to $Max = 2452885.3992 + 0.0942075 \times E$. In 2011 Wils et al. updated elements of V2455 Cyg to $Max = 2452885.399 + 0.094206008 \times E$. After that, only some times of maximum light of variable star V2455 Cyg were published. Peña et al. (2019) using $uvby - \beta$ photoelectric photometry, determined some physical characteristics of V2455 Cyg such as an effective temperature between $7200\, K$ and $7900\, K$ and absolute visual magnitude between $2.066\, mag$ and $1.075\, mag$.

In this paper, we presented new observations of V2455 Cyg, obtained times of maximum light and investigated the changes of the pulsation period. We also analyzed the pulsation frequency of V2455 Cyg and determined physical parameters of this δ Scuti star.

## 2 NEW V-BAND OBSERVATIONS

New photometric observations of V2455 Cyg were carried out at RIAAM[1] observatory ($\lambda = 37° 23' 58.095'' E, \phi = 46° 16' 9.019'' N$) on 2017 September 19 and 23, with 12-inch MeadeLX200

---

[1] Research Institute for Astronomy and Astrophysics of Maragha

S. Ostadnezhad, Gh. Forozani & M. Ghanaatian

**Table 1** The Coordinates and Magnitude of Variable, Comparison and Reference Stars.

| Stars | Name | RA | Dec | V (mag) | B (mag) |
|---|---|---|---|---|---|
| Variable | V2455 Cyg | $21^h28^m24.559^s$ | $+46°\,40'\,30.84"$ | 8.84 | 9.05 |
| Comparison | BD +46 3328 | $21^h28^m30.007^s$ | $+46°\,40'\,24.99"$ | 9.54 | 9.52 |
| Reference | TYC 3590-1667-1 | $21^h28^m08.164^s$ | $+46°\,37'\,59.18"$ | 12.86 | 12.50 |

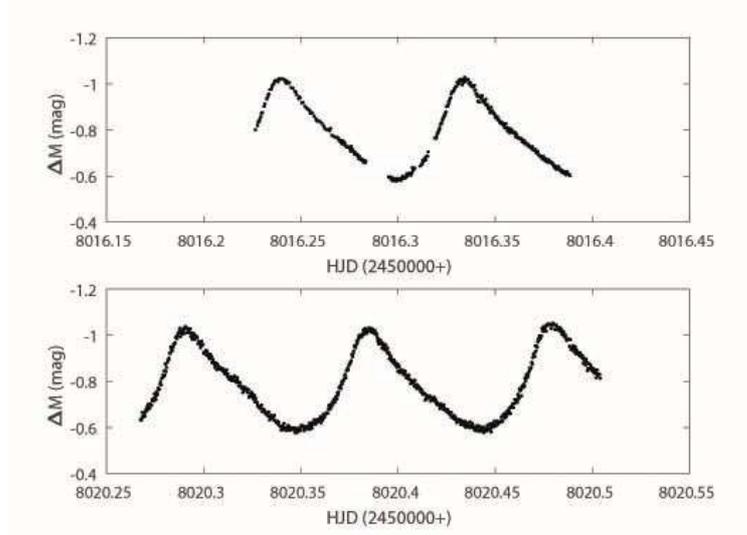

**Fig.** 1 The V-band observed light curves of V2455 Cyg in two nights.

Schmidt-Cassegrain Telescope equipped with a SBIG STX-16803 CCD camera. During the observations, V JohnsonCousins filter were used and the exposure time for each image was considered 20 s. A total of 1146 frames were obtained and 26 bias, dark and flat images were taken at each night. Data reduction was performed using MaxIm DL v5 software. BD +46 3328 and TYC 3590-1667-1 were used as the comparison and reference stars, respectively. Table 1 tabulates the coordinates and magnitudes of V2455 Cyg, comparison and reference stars which taken from SIMBAD astronomical database. New data (i.e. HJD and $\Delta M$) of V2455 Cyg during two nights observation are listed in Table 2 and new $V$-band light curves are shown in Fig. 1.

3 **TIMES OF MAXIMUM AND O-C CURVE ANALYSIS**

During two nights of new observations, five times of maximum light were observed (Fig. 1) and the timings of maxima were obtained by the 5th polynomial function fitting on each maximum light. These maxima and their errors are listed in Table 3. In order to investigation of the period of V2455 Cyg, we compiled 54 times of maximum light from different literatures together with new maxima. All light times of maxima are listed in the first and fifth columns and their references are tabulated in the fourth and eighth columns of Table 4. Utilizing all maxima, new linear ephemeris of V2455 Cyg was determined:

$$HJD_{max} = 2458016.237049(509) + 0.094206044^d(14) \times E \qquad (1)$$

and the ephemeris given by Wils et al. was improved:

$$HJD_{max} = 2452885.399056(429) + 0.094206044^d(14) \times E \qquad (2)$$

**Table 2** New V-Band Observations of V2455 Cyg.

| HJD | ΔM | HJD | ΔM |
|---|---|---|---|
| 2458016.226615 | -0.800 | 2458020.267553 | -0.633 |
| 2458016.227679 | -0.819 | 2458020.267854 | -0.631 |
| 2458016.228061 | -0.834 | 2458020.268166 | -0.640 |
| 2458016.228432 | -0.835 | 2458020.268479 | -0.664 |
| 2458016.228814 | -0.845 | 2458020.268791 | -0.648 |
| 2458016.229335 | -0.857 | 2458020.269104 | -0.647 |
| 2458016.229705 | -0.872 | 2458020.269416 | -0.658 |
| 2458016.230110 | -0.874 | 2458020.269729 | -0.650 |
| 2458016.231071 | -0.901 | 2458020.270041 | -0.670 |
| 2458016.231453 | -0.912 | 2458020.270354 | -0.677 |
| 2458016.232413 | -0.934 | 2458020.270666 | -0.671 |
| 2458016.232841 | -0.940 | 2458020.270979 | -0.691 |
| 2458016.233212 | -0.951 | 2458020.271291 | -0.700 |
| 2458016.233582 | -0.955 | 2458020.271592 | -0.679 |
| 2458016.234138 | -0.975 | 2458020.271905 | -0.685 |
| ... | ... | ... | ... |

**Table 3** New Light Maximum Times of V2455 Cyg.

| Date of observation | HJDmax | Error |
|---|---|---|
| 19 September 2017 | 2458016.240334 | 0.015814 |
| 19 September 2017 | 2458016.334760 | 0.026290 |
| 23 September 2017 | 2458020.291903 | 0.027661 |
| 23 September 2017 | 2458020.386033 | 0.029445 |
| 23 September 2017 | 2458020.480130 | 0.030144 |

Using the new ephemeris in Eq. 1, epochs and the O-C values were calculated and listed in Table 4. The O-C diagram versus $E$ is plotted in Fig. 2. As seen in Fig. 2, three maxima (shown as squares) at HJD 2456862.3963, 2456862.4903 and 2456862.5842 (Hubscher & Lehmann 2015) are too scattered rather than other times. So these times were discarded and the O-C curve was plotted in the upper panel of Fig. 3 as dots. The parabolic shape in the O-C curve (Fig. 3) shows increasing the pulsation period with time. By least squares method, the best parabolic function was fitted on the O-C curve (solid line in the upper panel of Fig. 3) and the quadratic ephemeris was obtained:

$$HJD_{max} = 2458016.239211(407) \\ + 0.094206204(30) \times E \\ + 2.417(469) \times 10^{-12} \times E^2 \quad (3)$$

with the standard deviation of σ = 0.00111 d. The quadratic term in Eq. 3 indicates that the period change rate of V2455 Cyg is dP/dt = 1.87(38) × $10^{-8}$ d yr$^{-1}$ and (1/P) dP/dt = 1.99(41) × $10^{-7}$ yr$^{-1}$. The lower panel of Fig. 3, shows the residuals from Eq. 3.

## 4 FREQUENCY ANALYSIS AND PHYSICAL PARAMETERS

Frequency analysis of V2455 Cyg was performed by using Period04 (Lenz & Breger 2005) software, which is based on the Fourier Transform,

$$f(t) = Z + \sum_i A_i \sin(2\pi(\Omega_i t + \Phi_i)) \quad (4)$$

on the observational data, where $\Omega_i$, $A_i$, $t$ and $\Phi_i$ are frequency, amplitude, time and phase, respectively. The results of the fitting (Eq. 4) on the new data (points in Fig. 4) are presented in Table 5 (i.e., frequency, amplitude and phase) and the synthetic light curves of this fitting are shown as solid lines in Fig. 4. As listed in the last column of Table 5, the frequencies which their signal-to-noise ratio (S/N) are greater than 4 were considered (Breger et al. 1993). The power spectrum of V2455 Cyg is shown in Fig. 5. Given the amplitude values listed in the second column of Table. 5 and the power spectrum in the Fig. 5, can be concluded that the fundamental frequency of V2455 Cyg is $10.61574 d^{-1}$.

**Table 4** All Times of Light Maxima of Pulsating Star V2455 Cyg.

| HJD 2450000+ | Epoch | O-C | Reference | HJD 2450000+ | Epoch | O-C | Reference |
|---|---|---|---|---|---|---|---|
| 2885.3992 | -54464 | 0.0001 | Wils et al. (2003) | 4759.3447 | -34572 | -0.0010 | Wils et al. (2009) |
| 2887.3778 | -54443 | 0.0004 | Wils et al. (2003) | 4759.4389 | -34571 | -0.0010 | Wils et al. (2009) |
| 2887.4720 | -54442 | 0.0004 | Wils et al. (2003) | 4759.5332 | -34570 | -0.0009 | Wils et al. (2009) |
| 2887.5657 | -54441 | -0.0001 | Wils et al. (2003) | 5365.4676 | -28138 | 0.0002 | Wils et al. (2011) |
| 2887.6599 | -54440 | -0.0001 | Wils et al. (2003) | 5365.5619 | -28137 | 0.0003 | Wils et al. (2011) |
| 2928.2634 | -54009 | 0.0006 | Wils et al. (2003) | 5373.4747 | -28053 | -0.0002 | Wils et al. (2011) |
| 2928.3582 | -54008 | 0.0012 | Wils et al. (2003) | 5417.5631 | -27585 | -0.0002 | Wils et al. (2011) |
| 2928.4520 | -54007 | 0.0008 | Wils et al. (2003) | 5820.3880 | -23309 | -0.0004 | Wils et al. (2012) |
| 2928.5465 | -54006 | 0.0011 | Wils et al. (2003) | 5834.5190 | -23159 | -0.0003 | Hubscher & Lehmann (2012) |
| 2929.2996 | -53998 | 0.0005 | Wils et al. (2003) | 5855.3390 | -22938 | 0.0002 | Wils et al. (2012) |
| 2929.3940 | -53997 | 0.0007 | Wils et al. (2003) | 6107.3410 | -20263 | 0.0010 | Wils et al. (2013) |
| 2929.4885 | -53996 | 0.0010 | Wils et al. (2003) | 6107.4350 | -20262 | 0.0008 | Wils et al. (2013) |
| 2929.5823 | -53995 | 0.0006 | Wils et al. (2003) | 6107.5290 | -20261 | 0.0006 | Wils et al. (2013) |
| 2931.4667 | -53975 | 0.0009 | Wils et al. (2003) | 6862.3963 | -12248 | -0.0051 | Hubscher & Lehmann (2015) |
| 4357.5561 | -38837 | -0.0008 | Wils et al. (2009) | 6862.4903 | -12247 | -0.0053 | Hubscher & Lehmann (2015) |
| 4642.4354 | -35813 | -0.0006 | Wils et al. (2009) | 6862.5842 | -12246 | -0.0056 | Hubscher & Lehmann (2015) |
| 4642.5296 | -35812 | -0.0006 | Wils et al. (2009) | 6867.4869 | -12194 | -0.0016 | Hubscher (2015) |
| 4646.4860 | -35770 | -0.0009 | Wils et al. (2009) | 6867.5832 | -12193 | 0.0004 | Hubscher (2015) |
| 4652.5157 | -35706 | -0.0003 | Wils et al. (2009) | 6914.5928 | -11694 | 0.0012 | Hubscher (2015) |
| 4694.4383 | -35261 | 0.0006 | Wils et al. (2009) | 7565.9307 | -4780 | -0.0015 | Peña et al. (2019) |
| 4694.5327 | -35260 | 0.0008 | Wils et al. (2009) | 7574.8801 | -4685 | -0.0016 | Peña et al. (2019) |
| 4730.3298 | -34880 | -0.0004 | Wils et al. (2009) | 7575.9157 | -4674 | -0.0023 | Peña et al. (2019) |
| 4730.4239 | -34879 | -0.0005 | Wils et al. (2009) | 8016.2403 | 0 | 0.0033 | This Study |
| 4730.5182 | -34878 | -0.0004 | Wils et al. (2009) | 8016.3348 | 1 | 0.0035 | This Study |
| 4730.6126 | -34877 | -0.0003 | Wils et al. (2009) | 8020.2919 | 43 | 0.0040 | This Study |
| 4758.4031 | -34582 | -0.0005 | Wils et al. (2009) | 8020.3860 | 44 | 0.0039 | This Study |
| 4758.4973 | -34581 | -0.0005 | Wils et al. (2009) | 8020.4801 | 45 | 0.0038 | This Study |

**Table 5** Results of Frequency Analysis of V2455 Cyg.

| $\Omega_i (d^{-1})$ | $A_i$ (mmag) | $\Phi_i (0-1)$ | S/N |
|---|---|---|---|
| 10.61574(28) | 192.440(685) | 0.369(1) | 66.3 |
| 21.22675(564) | 52.613(685) | 0.067(2) | 37.0 |
| 32.08961(379) | 21.701(685) | 0.781(5) | 14.4 |
| Zero point= −0.769355 mag | | | |
| Residuals = 0.0152 mag | | | |

In order to calculate the physical parameters, considering the temperature of V2455 Cyg which varies between $7200 K$ and $7900 K$ (Peña et al. 2019), mean temperature ($T = 7490 K$) and absolute visual magnitude ($M_V = 1.57\ mag$) values were used. According to Table of Flower, which the bolometric correction was calculated as a function of the effective temperature of the star, the bolometric correction is $BC = 0.035(1)\ mag$. Using the relation $M_{bol} = M_V + BC$, the bolometric absolute magnitude was determined to be $M_{bol} = 1.61(24)\ mag$. The surface gravity $g$ of $\delta$ Scuti V2455 Cyg was calculated from the following equation (Claret et al. 1990),

$$\log g = 2.68(\pm 0.10) - 1.21(\pm 0.11) \log P \tag{5}$$

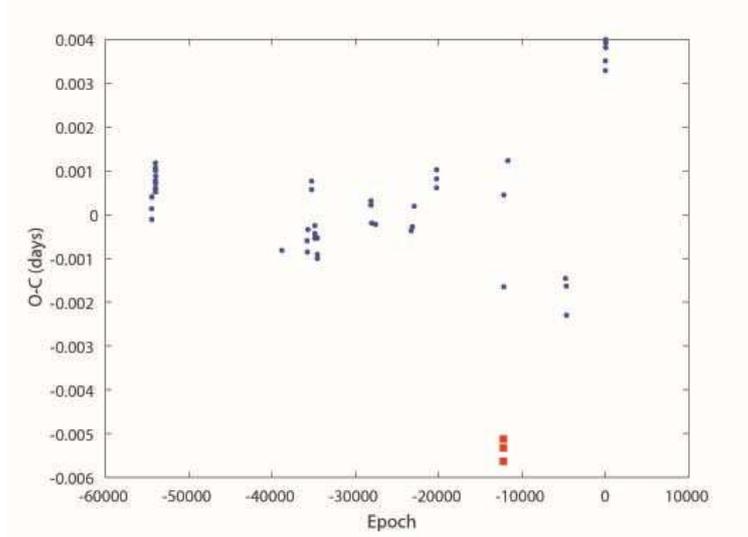

**Fig. 2** The O-C diagram of V2455 Cyg obtained of the new ephemeris (Eq. 1). The squares are maxima that are more scattered than other times.

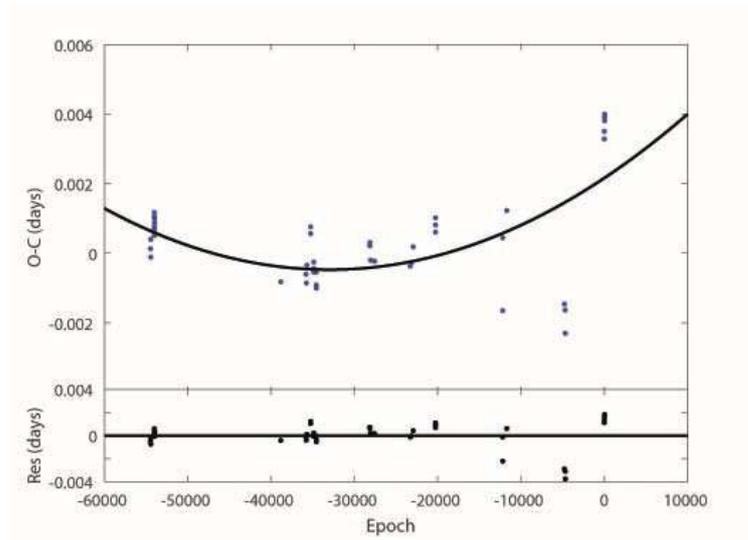

**Fig. 3** The O-C curve of V2455 Cyg based on 51 of maximum light. In the upper panel, the line is a parabolic fit (Eq. 3). The lower panel shows the residuals from parabolic fit.

where $P$ is the pulsation period. According to Eq. 5 the surface gravity will be $\log g = 3.92(10)$. Using the following relation (Breger et al. 1990),

$$\log R/R_\odot = -0.2 M_{bol} - 2 \log T + 8.472 \qquad (6)$$

which derived from the radiation law,

$$\log L/L_\odot = 2 \log R/R_\odot + 4 \log T/T_\odot \qquad (7)$$

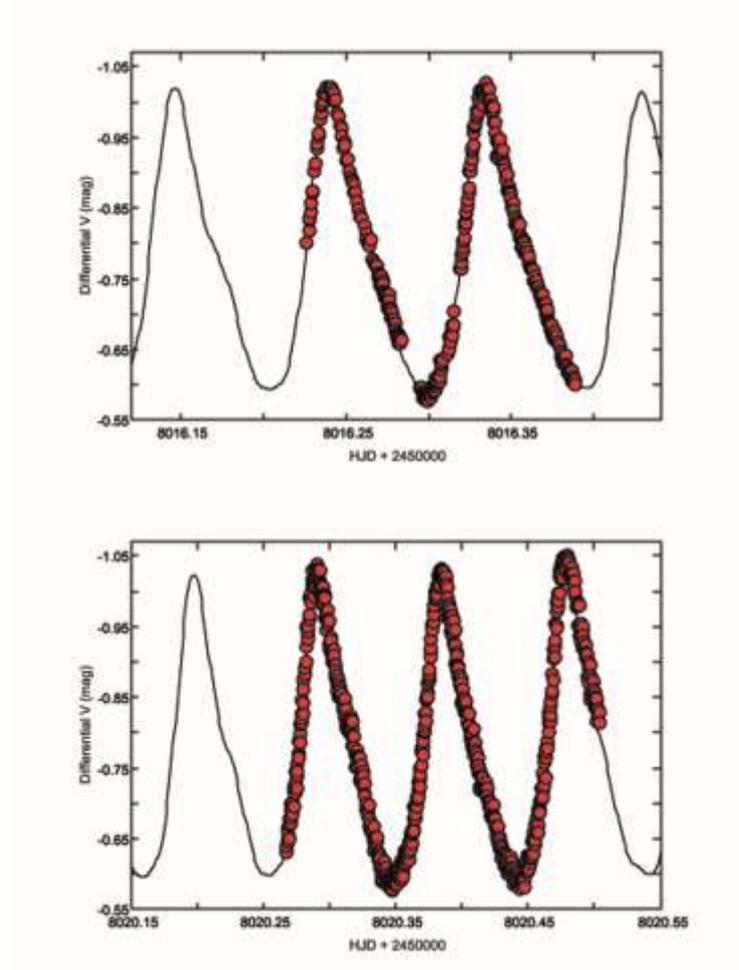

**Fig. 4** The V-band observed and synthetic light curves of V2455 Cyg. The points represent the observational data in two nights. The solid lines indicate the fit of three frequencies listed in Table 5.

the radius $R$ and luminosity $L$ of V2455 Cyg were obtained, where $T_\odot = 5770\,K$. Since the surface gravity is $g \sim M/R^2$, the mass of V2455 Cyg was determined from the following equation,

$$\log M/M_\odot = \log g - 4\log T/T_\odot + \log L/L_\odot - 4.44 \tag{8}$$

The values of physical parameters of V2455 Cyg at mean temperature are listed in Table 6. According to the period-mean density relation for pulsating stars (Petersen & Jørgensen 1972),

$$Q = P\sqrt{\bar{\rho}/\bar{\rho}_\odot} \tag{9}$$

where $\bar{\rho}$ and $\bar{\rho}_\odot$ are the mean density of the star and the Sun, the pulsation constant $Q$ is obtained from the following equation (Breger 1990b),

$$\log Q = -6.456 + \log P + 0.5 \log g + 0.1\, M_{bol} + \log T \tag{10}$$

Using obtained physical parameters and Eq. 10, the pulsation constant value of pulsating star V2455 Cyg will be $Q = 0.0327(21)\,d$.

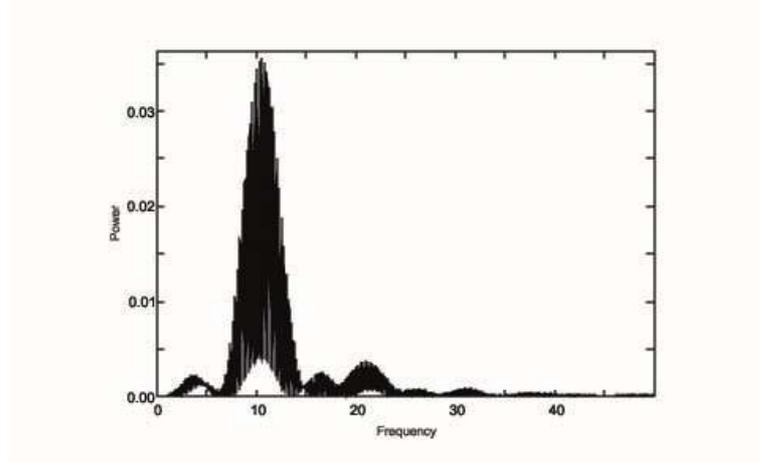

**Fig. 5** The power spectrum of V2455 Cyg.

**Table 6** The Physical Parameters of V2455 Cyg at Mean Temperature.

| Parameters | Value | Unit |
|---|---|---|
| Radius (R) | 2.52(19) | R$\odot$ |
| Mass (M) | 1.92(49) | M$\odot$ |
| Luminosity (L) | 18.12(3.58) | L$\odot$ |

## 5 DISCUSSION AND CONCLUSIONS

In this paper, we presented new $V$-band photometric observations of V2455 Cyg in two nights. During of observations, five times of maximum were obtained and new linear ephemeris (Eq. 1) was determined. By collecting all available maxima, the O-C diagram was plotted which indicates that the period of V2455 Cyg is increasing at the rate of $(1/P)\,dP/dt = 1.99 \times 10^{-7}\,yr^{-1}$. This slowly increasing period is in agreement with period changes in the majority of $\delta$ Scuti stars (Breger 1990a) such as GP And (Zhou & Jiang 2011) and YZ Boo (Yang et al. 2018) which their period is increasing.

We carried out the analysis of frequencies by using the Period04 software. Pulsation frequency analysis shows that the fundamental frequency of V2455 Cyg is 10.61574 $d^{-1}$. In the following, physical parameters of V2455 Cyg such as the mass and radius were obtained at mean temperature (Table 6) and the pulsation constant was calculated to be $Q = 0.0327\,d$. Since typical values of pulsation constants for the fundamental radial p modes in $\delta$ Scuti stars lie between $0.022 \leq Q \leq 0.033\,d$ (Breger & Bregman 1975), the value of pulsation constant shows that the variable star V2455 Cyg pulsates with the radial p-mode.

Fig. 6 shows the position of the $\delta$ Scuti stars in the Hertzsprung-Russell (Breger 1990a) which the population I $\delta$ Scuti stars (HADS), the population II $\delta$ Scuti stars (SX Phe) and $\delta$ Scuti V2455 Cyg are indicated as the plus, open circle and filled circle signs, respectively. As seen in Fig. 6, V2455 Cyg is situated above the zero age main sequence line which confirms that V2455 Cyg is a high amplitude $\delta$ Scuti star. For definite results about $\delta$ Scuti star V2455 Cyg, more photometric and spectroscopic observations are suggested.

**Acknowledgements** Authors was supported by Department of Physics, Payame Noor University Tehran Iran. We appreciate the Research Institute for Astronomy and Astrophysics of Maragha (RIAAM) for their cooperation.

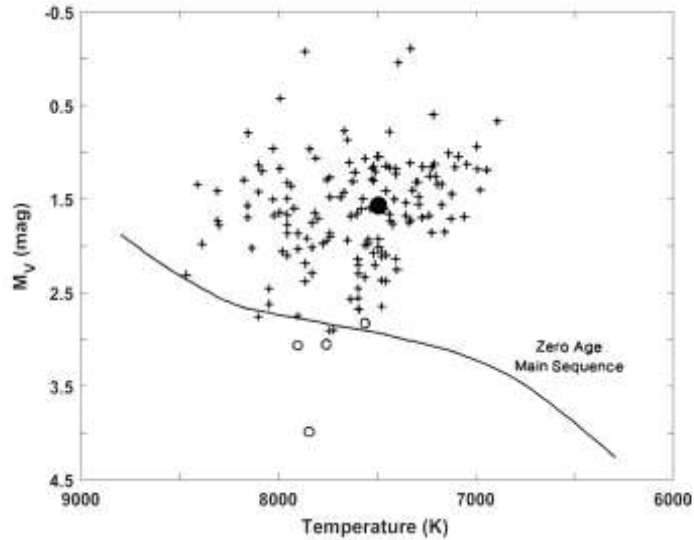

**Fig. 6** The position of V2455 Cyg (filled circle) in the H-R diagram at mean temperature. The plus signs indicate HADS stars and the open circles show SX Phe stars (Breger 1990a).